\documentstyle[11pt,paspconf,epsf,psfig]{article}

\newcommand{\Ha}{H$\alpha$}

\newcommand{\Hb}{H$\beta$}
\newcommand{\Ms}{$M_{\odot}$}

\newcommand{\lya}{Ly$\alpha$}

\def\ltsima{$\; \buildrel < \over \sim \;$}
\def\simlt{\lower.5ex\hbox{\ltsima}}
\def\gtsima{$\; \buildrel > \over \sim \;$}
\def\simgt{\lower.5ex\hbox{\gtsima}}

\begin{document}

\title{The Discovery of Primeval Galaxies and 
the Epoch of Galaxy Formation}

\author{Max Pettini}
\affil{Royal Greenwich Observatory, Madingley Road, Cambridge CB3 0EZ,
England}

\author{Charles C. Steidel, Kurt L. Adelberger, Melinda Kellogg}
\affil{Palomar Observatory, Caltech 105$-$24, Pasadena, CA 91125, USA}

\author{Mark Dickinson}
\affil{Department of Physics and Astronomy, The Johns Hopkins University,
Baltimore, MD 21218, USA}

\author{Mauro Giavalisco}
\affil{The Carnegie Observatories, 813 Santa Barbara Street, Pasadena,
CA 91101, USA}


\begin{abstract}

In the last two years there have been major advances in our
ability to identify and study normal star forming galaxies at
high redshifts, when the universe was only 15\% of its present
age. We review the steps which have led to the discovery of a
widespread population of objects at $z \sim 3$ with many of the
characteristics which we expect for primeval galaxies, and 
emphasize in particular the advantages of a colour selection
technique which targets the Lyman discontinuity at 912~\AA.

Star forming galaxies at $z = 3$ resemble local starbursts,
although they are typically more luminous by more than one order
of magnitude. The ultraviolet continuum is dominated by the
integrated light of O and early B type stars and shows prominent
interstellar absorption lines which are often blueshifted
relative to the systemic velocity of the galaxy, indicating
highly energetic outflows in the interstellar medium. \lya\
emission is generally weak, probably as a result of resonant
scattering. The spectral slope of the ultraviolet continuum and
the strength of the \Hb\ emission line, which we have detected in
a few cases with pilot observations in the infrared $K$ band,
suggest that some interstellar dust is already present in these
young galaxies and that it attenuates their UV luminosities by a
factor of $\sim 3$. 

The efficiency of our photometric selection technique has allowed
us to establish that large scale concentrations of galaxies were
already in place at $z = 3$; these structures may be the
precursors of today's rich clusters of galaxies, at a time when
they were beginning to decouple from the Hubble expansion. In the
context of Cold Dark Matter models of structure formation, the
galaxies we see must be associated with very large halos, of mass
$M \simgt 10^{12}$\Ms, in order to have developed such strong
clustering at $z = 3$. We conclude by pointing out the need for 
infrared space observatories, such as the proposed {\it Next
Generation Space Telescope}, for pushing the quest for the origin
of galaxies beyond $z = 5$.

\end{abstract}

\section{Introduction}

The quest for the origin of galaxies has been one of the main themes of 
observational cosmology for many years. A key aspect of this search is the 
identification of `primeval' galaxies---a somewhat lose concept given that 
we don't know how galaxies form, but generally taken to mean the 
progenitors of galaxies like the Milky Way at the time when they first 
assembled a significant fraction of their mass and began forming their 
first generations of stars.  
Until recently the search for primeval galaxies had been a highly 
frustrating affair with only a handful of objects, mostly discovered 
serendipitously or through gravitational lensing, as the meagre return for 
the investment of many months of observing time on large telescopes.

This, we now realize, was due less to a lack of adequate instrumentation 
than to the fact that we did not know how to recognise what we were 
looking for.
A typical deep CCD image obtained at the prime focus of a 4 m telescope
shows some 3000 galaxies which are at distances stretching from the 
vicinity of the Milky Way to the most distant reaches of the universe, 
corresponding to look-back times of 90\% of the age of the universe.
Thus, such a CCD image in principle contains 
much of the information required to 
identify the origin of galaxies and follow their evolution over most of 
the Hubble time. 
The challenge until now has been to devise an efficient way to sort this 
multitude of galaxies according to their redshifts and ages and 
thereby identify the young counterparts 
of present day luminous galaxies.

The situation has improved dramatically in 
the last two years and objects which conform 
closely to our ideas of a primeval galaxy are now being discovered 
routinely and in large numbers. 
In this conference contribution
we give an account of these recent exciting developments.
We first describe the method 
which has proved to be most profitable for identifying star forming
galaxies at high redshifts. 
We then review some of the most significant properties
of this population of objects, deduced  
from the analysis of their redshift distribution and their spectra.
We end with some comments on future prospects, focussing in 
particular on the infrared spectral region and on the key role which the 
{\it Next Generation Space Telescope} will play in 
pushing this field of research to the earliest epochs.

Before proceeding just one word of clarification. 
When dealing with distant galaxies astronomers normally
refer to their redshift, because this 
is the quantity which is directly measured from the spectra.
Redshift, however, is only a proxy for 
look-back time---we are interested in 
how far in the past we observe
a particular galaxy at redshift $z$. 
The mapping of redshift to look-back time is not yet as accurate as we 
would like, although the precision in the measurement
of the Hubble constant
has improved in the last few years.
For reference, Table 1 gives the look-back times
corresponding to values of redshift used most often in this review, 
adopting a Hubble constant 
$H_0 = 70$~km~s$^{-1}$~Mpc$^{-1}$ and 
a deceleration parameter for the expansion of the universe
$q_0 = 0.1$
(unless otherwise stated, these values are assumed throughout 
this article).

\begin{table}
\caption{Look-Back Time as a Function of Redshift.}\label{tbl-1}
\begin{tabular}{rrr}
\\
$z$ & ~~$T$ (Gyr)\tablenotemark{a} & ~~$T/T_{\infty}$\tablenotemark{a}\\
\tableline
\\
0 & 0 & 0 \\
0.5 & 4.6 & 0.39 \\
1 & 6.8 & 0.57 \\
2 & 8.8 & 0.74 \\
3 & 9.8 & 0.83 \\
4 & 10.3 & 0.87 \\  
10 & 11.4 & 0.96 \\ 
$\infty$ & 11.9 & 1.00\\
\end{tabular}
 
\tablenotetext{a}{$H_0 = 70$~km~s$^{-1}$~Mpc$^{-1}$; $q_0 = 0.1$}

\end{table}

\section{Imaging in the Ultraviolet Stellar Continuum}

The galaxies we see around us today exhibit a wide variety of shapes and 
sizes. 
Approximately half the stars in the present day universe are found 
in `old' systems, elliptical galaxies and the bulges of spirals, which are 
collectively referred to as the spheroidal component of the galaxy 
population. The conventional view has been that these galaxies 
formed a long time ago, probably before $z = 1$, and relatively 
rapidly, whereas the assembly of the more fragile 
disks of spiral galaxies such as the 
Milky Way was presumably a slower and more protracted process.
In this scenario one would expect the formation of ellipticals and bulges 
to be a period of intense star formation activity.

Star forming galaxies in our vicinity have very characteristic 
visible spectra.  Between 3500 and 7500 \AA\ the spectra are 
dominated by strong, narrow emission lines of hydrogen, 
oxygen and nitrogen produced in the nebulae associated with newly formed 
massive stars. Compared with these strong emission lines the underlying 
blue continuum radiation from the stars themselves is relatively faint.
At redshift $z = 3$ however, the portion of the spectrum recorded by an 
optical telescope is the rest-frame ultraviolet, between 1000 and 2000 \AA.
It was only natural that the first searches for galaxies at high 
redshift should have targeted, either spectroscopically or by 
narrow-band imaging, what was expected to be the strongest nebular
emission line in the ultraviolet, the \lya\ line of neutral 
hydrogen at a rest wavelength of 1215.67~\AA.
Despite massive efforts, searches for \lya\ emission at high redshift 
have generally been disappointing 
(e.g. Thompson, Djorgovski, \& Trauger 1995).
It is only with the benefit of hindsight that we now fully appreciate
how difficult it is for \lya\ photons to escape the galaxy where they are
produced, given the large cross-section for absorption and subsequent 
scattering by H~{\sc I} atoms (Meier \& Terlevich 1981;
Neufeld 1991; Charlot \& Fall 1993; 
Valls-Gabaud 1993).

%
%
\begin{figure}
\hspace*{-0.1in}
\psfig{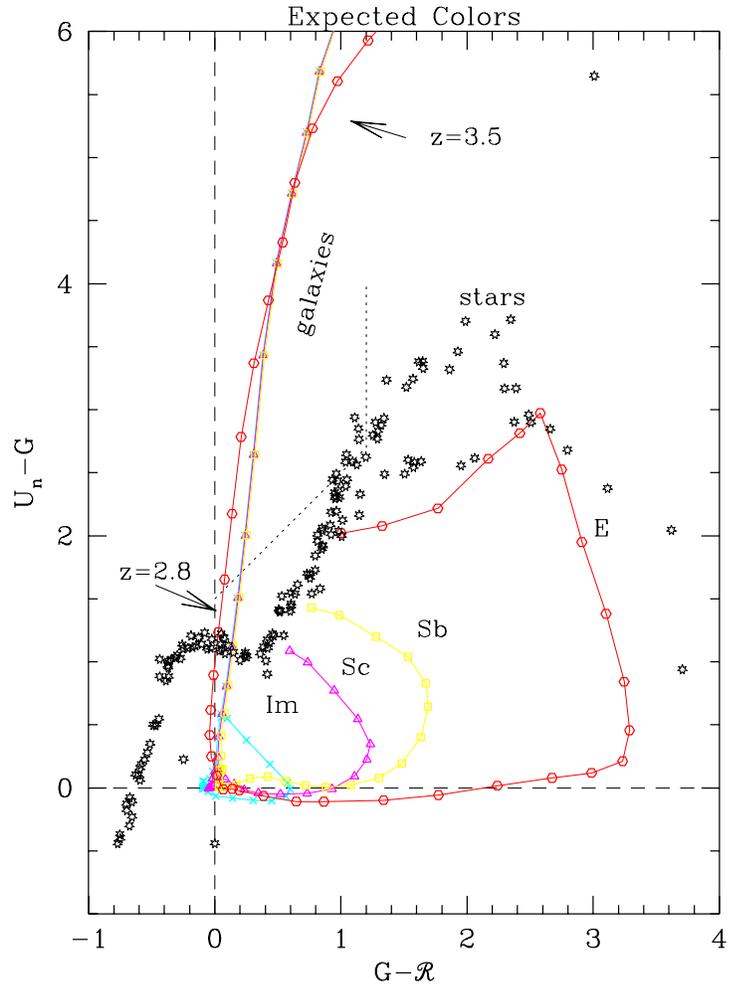}
\hspace*{-0.75in}
\caption{
Colour evolution with redshift 
of galaxies of different spectroscopic type 
in the $U_n$, $G$, ${\cal R}$ filter systems devised by 
Steidel \& Hamilton (1993).
Magnitudes are defined in the AB system, so that a galaxy with a flat 
spectrum in $f_{\nu}$ 
(where $f_{\nu}$ is the flux per frequency unit)
has $(U_n - G)$ = $(G- {\cal R})$ = 0.0\,.
The curves start at $z = 0$ and each point along a track corresponds to 
a redshift increment $\Delta z  = 0.1$\,.
Colours were calculated from the model spectral energy distributions by 
Bruzual \& Charlot (1996) and Madau's (1995) estimate of the opacity due 
to \lya\ forest and Lyman continuum absorption by 
intervening gas. 
The dotted lines indicate the boundary of our photometric selection criteria 
for robust candidates. 
The star symbols show the colours of Galactic 
stars of different spectral type.
}
\end{figure}

The method which has turned out to be successful in finding 
normal galaxies at $z = 3$ was developed by
Steidel \& Hamilton (1992, 1993) and is based on the realization that 
the limit of the Lyman series near 912~\AA, the wavelength of photons 
with sufficient energy to ionise hydrogen, produces an obvious
discontinuity in the far-UV spectrum of {\it any} 
star forming galaxy.
This Lyman `break' has a three-fold origin: the intrinsic 
drop in the spectra of hot O and B type stars which dominate the
integrated spectrum at 
ultraviolet wavelengths (e.g. Cassinelli et al. 1995);
the absorption by the neutral interstellar medium within a star forming 
galaxy (Leitherer et al. 1995a; Gonz\'{a}lez Delgado et al. 1997); and 
the opacity of the intervening intergalactic medium.
This last effect, which has been quantified by Madau (1995)
from the well known statistics of QSO absorption line spectra,
turns out to be the overriding 
factor which determines the colour of high redshift galaxies. 

Steidel \& Hamilton devised a three filter system optimised for 
selecting Lyman break galaxies. Two of the filters,
$U_n$ and $G$ in their notation, have passbands 
respectively below and above the Lyman limit at $z = 3$, 
while the third filter, 
${\cal R}$, is further to the red.
In deep images of the sky taken through these three filters 
$z = 3$ galaxies are readily distinguished from the bulk of 
lower redshift objects  
by their red $(U_n - G)$ and blue $(G - {\cal R})$ colours.  
As can be seen from Figure 1,  
high redshift galaxies fall in a well-defined region of the 
colour-colour plot, irrespective of their spectroscopic type.
In the survey which followed the initial observations by 
Steidel \& Hamilton (1992), we designated objects with
$(G - {\cal R}) < 1.2$ and $(U_n - G) > 1.5 + (G - {\cal R})$
(that is with colours within the dotted lines in Figure 1)
as `robust' candidates for high redshift galaxies.
In order to explore the boundary for high-$z$ galaxies in the two colour plane
we also selected a number of `marginal' candidates, defined to be objects 
with $(G - {\cal R}) < 1.2$ and 
$1.0 + (G - {\cal R}) < (U_n - G) < 1.5 + (G - {\cal R})$, that is
objects falling in a strip half a magnitude wide in 
$(U_n - G)$ below the sloping line in Figure 1.

%
%
\begin{figure}
\vspace*{+0.80in}
\hspace*{+0.7in}
\psfig{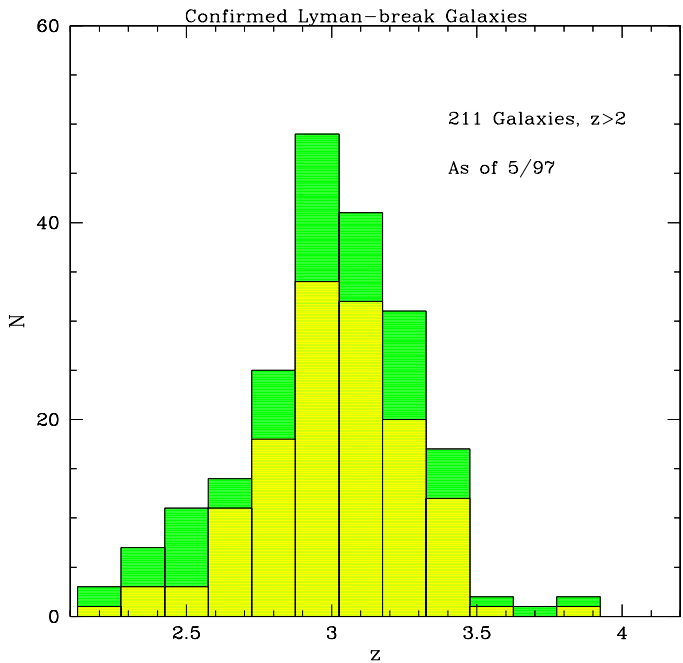}
\hspace*{-0.75in}
\vspace*{-0.85in} 
\caption{Redshift distribution of spectroscopically confirmed Lyman break 
galaxies in our survey. The light area refers to the robust candidates only,
while the darker area shows the sum of robust and marginal
candidates (the corresponding
photometric selection criteria are explained in the text).
}
\end{figure}

Follow-up spectroscopy with the Keck telescopes has confirmed
the efficiency of this photometric selection 
technique (Steidel et al. 1996). 
Approximately 95\% of the robust candidates are indeed high redshift 
galaxies, the remainder being mainly faint Galactic stars.
At the time of this conference (May 1997) we have imaged  
approximately 800 square arcminutes of sky in a dozen fields.
Some of the fields are centred on high redshift QSOs with known
intervening Lyman limit systems, while others are random pointings 
in areas which in some cases have also been imaged with the 
{\it Hubble Space Telescope}.
At a limiting magnitude ${\cal R} \leq 25$, which roughly corresponds to the 
magnitude of a galaxy like the Milky Way (an $L^{\ast}$ galaxy)
at $z = 3$, the density of Lyman break galaxies 
on the plane of the sky is
$\approx 0.5$~arcmin$^{-2}$ (Steidel, Pettini, \& Hamilton 1995),
only $\approx 1\%$ of the surface density 
of {\it all} galaxies 
brighter than this limit.
Even so, a deep $2048 \times 2048$ pixel CCD image recorded at the prime 
focus of the 5 m Hale telescope at Palomar
will yield $\sim 40$ robust candidates, a vast improvement over the 
dearth of {\it bona fide} high redshift galaxies which persisted until a few 
years ago!

Figure 2 shows the redshift histogram of all the spectroscopically 
confirmed galaxies in our sample. Clearly, the 
$U_n$, $G$, ${\cal R}$ filter system is most effective at selecting 
galaxies at redshifts $2.6 < z < 3.4$.
As described by Piero Madau (see Figure 4 of his article in this volume)
the same Lyman break technique has been applied to the
$U,B,V,I$ images of the {\it Hubble Deep Field}
to identify galaxies at redshifts 
$2.0 < z < 3.4$ (the $U$ drop-outs) and $3.5 < z < 4.5$ 
($B$ drop-outs).

\section{Large Scale Concentrations of Matter at High Redshift}

An immediate bonus of an efficient photometric selection method
is the ability to probe with the minimum of effort
the three-dimensional distribution of the objects of interest.
For Lyman break galaxies at $z \simeq 3$ there is the added advantage 
that the angular scales sampled map to 
comoving scales which are sufficiently large to be of interest. 
This is generally not the case for `pencil beam' surveys 
at lower redshifts which essentially give a one-dimensional view of 
clustering along the line of sight. 
The anisotropy of the cosmic microwave background tells us that
seed fluctuations were present at the earliest times; the extent to which 
these have grown after the first 15\% of the age of the universe
depends sensitively on the cosmological parameters and 
can therefore be used to discriminate between different world models.
    
The first application of the Lyman break technique to the study of 
large scale structure at high redshift has produced very exciting 
results and led to the discovery of a large concentration of galaxies at 
$z = 3.090$ in one of our fields (see Figure 3). 
Our survey suggests that
such `spikes' in redshift space 
were already common at these epochs and
may well be examples of today's rich clusters of 
galaxies caught early in their evolution, when they were beginning
to break away 
from the Hubble expansion.

%
%
\begin{figure}
\hspace*{+0.55in}
\psfig{figure=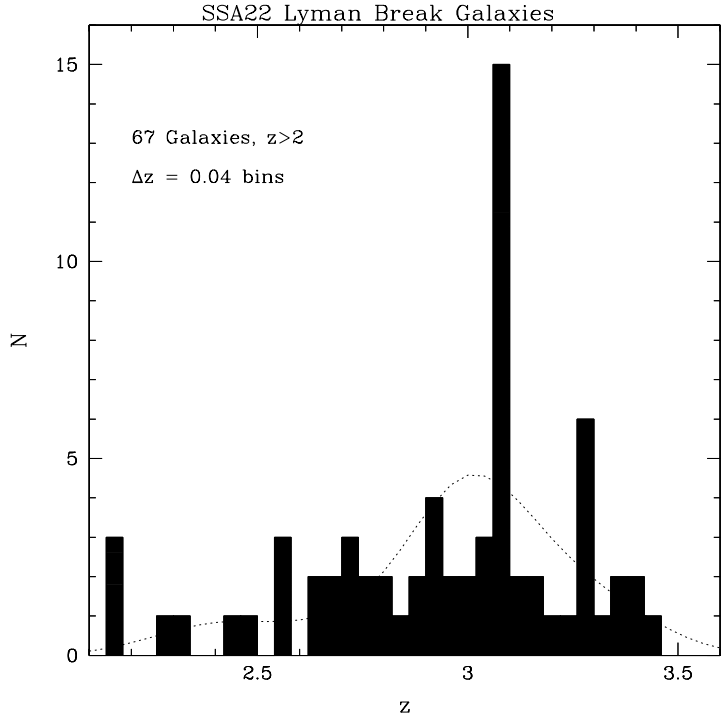,width=10.0cm,height=10.0cm,angle=0.0}
\caption{Redshift histogram of all spectroscopically confirmed 
galaxies in a 9 by 18 arcminute strip of sky near the Hawaii deep field 
SSA22. The dotted line shows the selection function for our entire survey
(normalised to the total number of galaxies in the histogram). 
The prominent spike at mean redshift 
$\langle z \rangle = 3.090$ is significant at the 99.8\% confidence level.
}
\end{figure}
%
%
%

%
%
\begin{figure}
\hspace*{+0.280in}
\psfig{figure=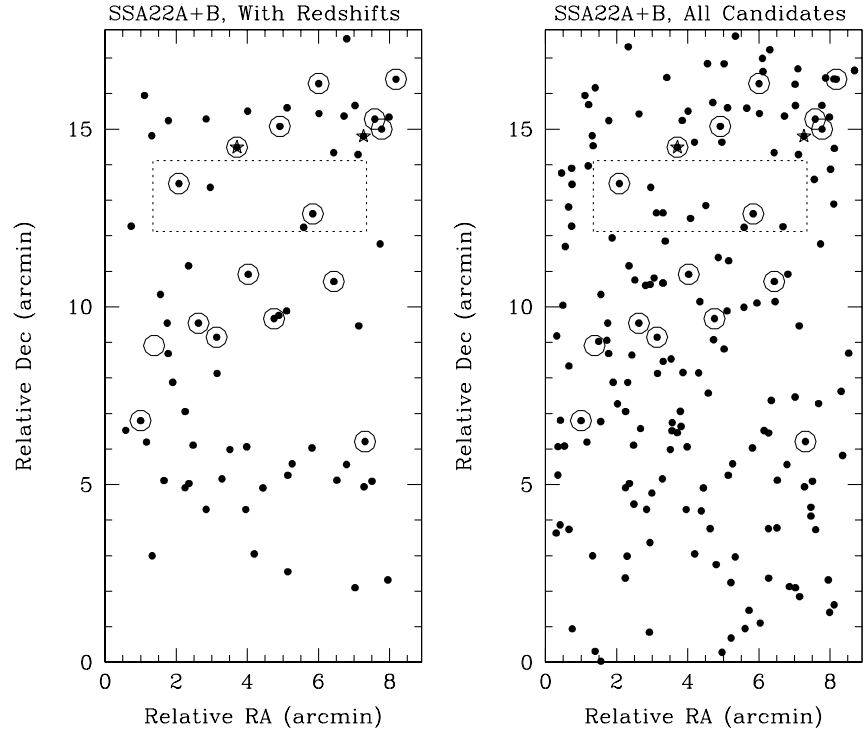,width=12.0cm,height=12.0cm,angle=0.0}
\caption{{\it Left panel:} Distribution on the plane of the sky of all 
spectroscopically confirmed 
Lyman break objects with redshift $z > 2$
in the $9 \times 18$~arcminute strip. The 16 objects with
$\langle z \rangle = 3.090 \pm 0.02$ are circled
(the empty circle refers to an object discovered serendipitously
from \lya\ emission).
The star symbols indicate 
the two QSOs we found in this field.  
{\it Right panel:} All Lyman break candidates 
in the same area, selected according to both robust 
and marginal photometric criteria given in the text. 
The dotted region in both panels shows the 
Hawaii deep field SSA22 surveyed by Cowie et al. (1996).
}
\end{figure}

One of our best observed regions of sky consists of two adjacent 
$9 \times 9$~arcminute fields which we have imaged very deeply
(to a limiting $1 \sigma$ surface brightness of 29~mag~arcsec$^{-2}$
in all three $U_n$, $G$, and ${\cal R}$ passbands) 
with the COSMIC prime focus camera of the 5~m Hale telescope at Palomar 
Observatory. The strip includes one of the Hawaii deep fields designated 
SSA22 by Cowie et al. (1996) and overlaps with one of the 
Canada-France redshift survey fields (Lilly et al. 1995a).
We find a total of 181 Lyman break candidates (robust and marginal) 
brighter than ${\cal R} = 25.5$ in this area; their distribution on the 
plane of the sky is shown in the right-hand panel of Figure 4.
Subsequent Keck spectroscopy of a subset of the candidates
confirmed that 67 of them are at $z > 2$; of these, 15 galaxies and 1 QSO 
fall within a redshift interval $\Delta z = 0.04$ centred at 
$\langle z \rangle = 3.090$\,. 
The comoving dimensions of this structure 
in an Einstein-de Sitter universe ($\Omega_M = 1$)
are at least $14 h^{-1}_{70}$ by $10 h^{-1}_{70}$~Mpc in the
transverse direction and $16 h^{-1}_{70}$~Mpc along the line of sight.
On the basis of our sampling 
frequency we estimate that 
the concentration contains in excess of  $30 - 50$ galaxies 
brighter than $L^{\ast}$.

The implications of having found such a strong clustering signal
at $z = 3$ are considered in detail by Steidel et al. (1997) 
in the context of Cold Dark Matter theories for the growth of structure
in the universe 
(see David Weinberg's review at this conference).
After correcting for the effect of peculiar velocities, the
spike in the redshift histogram corresponds to an overdensity 
$\delta_r \simeq 2$ relative to the background distribution of mass 
on the scale of the structure ($\delta_r$ is the fractional 
excess of galaxies which an observer in one of the galaxies in the spike 
would measure relative to the average of many such volumes).
The probability of finding such an overdensity approaches unity only
if galaxies are very biased tracers of mass; we deduce
values of the bias parameter 
$b \equiv \delta_{gal}/\delta_{mass} \simeq 2$ for 
$\Omega_M = 0.2$ and as high as $b \sim 6$ if $\Omega_M = 1$.

In CDM models of galaxy formation (e.g. Baugh et al. 1997)
the first galaxies do indeed form in the highest density peaks 
and a high bias 
parameter is not unexpected at $z = 3$\,.
Since the more massive halos are more heavily biased (Mo \& White 1996),
the characteristic mass associated with the Lyman break galaxies can be 
deduced from the observed bias parameter; 
the above values of $b$ point to dark halo masses
$M \simgt 10^{12}$~\Ms. This is strong evidence in support of the view 
that in the Lyman break objects we see the progenitors of today's 
luminous galaxies ($L^{\ast}$ and brighter), rather than smaller fragments
undergoing an intense and short-lived burst of star formation. 
 
One of the two QSOs we have identified in our observations of the SSA22,
at redshift $z_{em} = 3.356$, lies {\it behind}
the concentration of galaxies giving rise to the spike (see Figure 4).
It is very interesting to find that the spectrum of this QSO shows 
absorption systems at the redshifts of the three most significant 
features in the histogram in Figure 3, $z = 2.93$, 3.09, and 3.28\,.
The implication seems to be that diffuse gas 
with near-unity  covering factor is associated with the structures we see 
in the large scale distribution of galaxies.
The combination of absorption and emission techniques will 
be very powerful for studying
the high redshift progenitors of rich clusters of galaxies.

\section{The Spectra of High Redshift Galaxies}

The sample of more than 200 ultraviolet spectra 
which we have assembled in our survey is a rich source of information on 
the physical properties of star forming galaxies at high redshift.
The spectra show many similarities with 
those of nearby starburst galaxies; accordingly our analysis  
draws extensively on the detailed studies of local starbursts 
carried out with the {\it HST} 
and described in Tim Heckman's article in this volume.
Here we focus in particular 
on: {\it (a)} the UV luminosities and implied star 
formation rates; {\it (b)} evidence for the presence of dust and the 
corresponding UV extinction; and  
{\it (c)} \lya\ emission and large-scale velocity fields in the 
interstellar medium.
The signal-to-noise ratio of most of our spectra, although relatively 
modest, is nevertheless adequate for addressing all three topics.
In addition, we have begun to study in detail a number of individual 
objects (generally among the more luminous in the sample) to which we 
have devoted long 
exposure times at moderately high spectral resolution.
The best case is  
the $z = 2.723$ galaxy 1512-cB58 (Yee et al. 1996).
This spectrum (reproduced in Figure 5) 
is of comparable quality with the best {\it HST} spectra of local 
starburst galaxies and is a veritable mine of information.
It now seems highly likely that cB58, which is
$\sim 4$ magnitudes brighter than the typical $z = 3$ galaxy
in our survey, 
is not extraordinarily luminous but gravitationally lensed
(Williams \& Lewis 1997; Seitz et al. 1997) and 
therefore presumably provides us with an unusually clear view of a normal
galaxy at high redshift.

\subsection{Ultraviolet Luminosities}

The typical high-$z$ galaxy in our survey, with 
${\cal R} = 24.5$, ($G-\cal{R}$) = 0.5, and $z = 3$,
has a far-UV luminosity
$L_{1500} = 1.3 \times 10^{41}~h_{70}^{-2}$
erg~s$^{-1}$~\AA$^{-1}$ at 
1500~\AA. 
It is instructive to compare this value with those measured in 
nearby starbursts. 
The UV luminosities we find at $z = 3$ are
$\approx 800$ times greater than that of the brightest star cluster
in the irregular galaxy NGC~4214 studied with {\it HST} by 
Leitherer et al. (1996),
and exceed by a factor of $\approx 30$ that of the most luminous local 
example, the Wolf-Rayet galaxy NGC~1741 
which contains $\approx 10^4$ O type stars
(Conti, Leitherer, \& Vacca 1996).

%
%
\begin{figure}[t]
\unitlength1cm
\begin{picture}(16.5,16.5)  
{\epsfxsize=16.5cm
\epsfbox{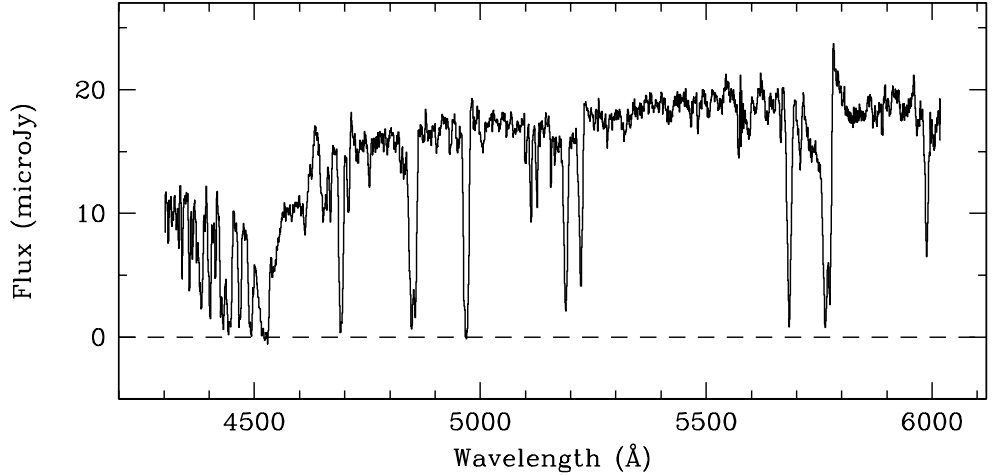}}
\end{picture}
\vskip -9cm
\caption{Spectrum of the $V = 20.64$, $z = 2.723$  
galaxy 1512-cB58 obtained with LRIS on the
Keck~I telescope in May and August 1996. With a total exposure time  
of 11\,400~s we reached S/N $\simeq 50$ per pixel at a resolution of 3.5 
\AA.}
\end{figure}

The ultraviolet spectra, in which we see the integrated 
continuum of O and early B stars, 
can in principle be used to estimate the star formation rate in 
a more direct way than the Balmer lines, which are produced
by the reprocessed ionizing radiation of 
the stars at the very tip of the IMF.
Adopting a continuous star formation model with an age 
greater than $10^8$~years 
and a Salpeter IMF from 0.1 to 100~\Ms\
(Bruzual \& Charlot 1996; Leitherer, Robert, \& Heckman 1995b),
the typical $L_{1500} = 1.3 \times 10^{41}~h_{70}^{-2}$
erg~s$^{-1}$~\AA$^{-1}$ corresponds to 
a star formation rate 
{\it SFR} $\simeq 8~h_{70}^{-2}$~\Ms~yr$^{-1}$.
This is probably a lower limit, since dust extinction
(see below) and a lower age would both raise this value
(for an age of $10^7$~years the implied  
{\it SFR} is greater by a factor of $\approx 1.7$).

It is interesting to note that even the brightest objects in our sample
fall well within the surface brightness distribution of local starbursts.
The highest values of $L_{1500}$ which we have found 
are $\sim 4-5$ times higher 
than the median. 
Adopting a typical extinction correction of a factor of 
$ \approx 3$ at 1500~\AA\ (see below)
and a typical half-light radius
$r \simeq 2$~kpc (Giavalisco, Steidel, \& Macchetto 1996),
we arrive at a star formation intensity
({\it SFR} per unit area) 
$\dot{\Sigma} \sim 13$~\Ms~yr$^{-1}$~kpc$^{-2}$\,.
This value compares well with the upper envelope of 
the ultraviolet sample 
considered by Meurer et al. (1997). 
Thus, the star forming galaxies which we are finding at high 
redshift appear to be spatially more extended versions
of the local starburst phenomenon. 
The same physical processes which 
limit the maximum star formation intensity in nearby starbursts, as 
discussed by Meurer et al., also seem to be at play  
in young galaxies at high redshift
which may well be undergoing their first episodes of star formation.

\subsection{Dust Extinction}

Dust is a ubiquitous component of the interstellar medium; 
given that galaxies at $z = 3$
are obviously already enriched in heavy elements, 
it is likely that some dust is mixed with the gas and stars
we observe.
Unfortunately, even relatively small amounts of dust can have a
significant effect in the rest-frame ultraviolet and thereby alter 
our view of high-$z$ galaxies. 
In particular, dust will: 
{\it (a)} extinguish  
resonantly scattered emission lines, most notably \lya;
{\it (b)} attenuate the UV continuum leading to underestimates of the
star formation rate; and
{\it (c)} redden the broad spectral energy distribution 
so that it resembles that of an older stellar population.
To some extent we have to learn to live with these
problems because of the inherent uncertainties of any dust corrections 
which arise mostly from the unknown shape of the extinction 
law.

The intrinsic slope of the integrated UV continuum 
of a star forming galaxy is a robust quantity which, as explained for 
example by Calzetti (1997a), varies little with 
the exact shape of the IMF or the age of the starburst.
Spectral synthesis models show that 
the continuum between 1200 and 1800 \AA\ is well 
approximated by a power law of the form $f_{\nu} \propto \lambda^{\alpha}$
with $\alpha$ between  $-0.5$ and 0. Similarly,
the empirical template starburst spectrum constructed by Calzetti (1997b)
has $\alpha = -0.1$\,.
In contrast, the galaxies we observe generally have UV spectral slopes
between 0 and $+1.5$; 
the spectrum of cB58 reproduced in Figure 5, for example, can be clearly 
seen to be redder than flat spectrum and  
has $\alpha = +1.3$\,.
Such (relatively) red spectra can result from an aging starburst or from 
an IMF lacking in massive stars, but we regard both possibilities as 
unlikely because with sufficiently high S/N 
we see directly the spectral signatures of O stars.
The most straightforward interpretation, in analogy with local 
starbursts, is that the spectra are reddened by dust extinction.

In Figure 6, reproduced from Dickinson et al. (1997),
we use the ($G-\cal{R}$) colours of the entire sample
of spectroscopically confirmed galaxies 
to estimate dust corrections to the star 
formation rates at high redshift. 
Assuming an intrinsic spectral slope 
$\alpha = -0.13$, as is the case for a   
Bruzual \& Charlot (1996) model
with 1~Gyr old continuous star formation and Salpeter IMF,
the curves labelled with different values of 
$E(B-V)$ at the top of the figure show the predicted 
($G-\cal{R}$) colour as a function of redshift, if the spectra of high-$z$ 
galaxies are reddened with an extinction law 
similar to that which applies to  
stars in the Small Magellanic Cloud.
The curves all rise to redder ($G-\cal{R}$) colour
with redshift due to the increasing line blanketing by the 
\lya\ forest (Madau 1995).

%
%
\begin{figure}
\hspace*{-0.25in}
\psfig{figure=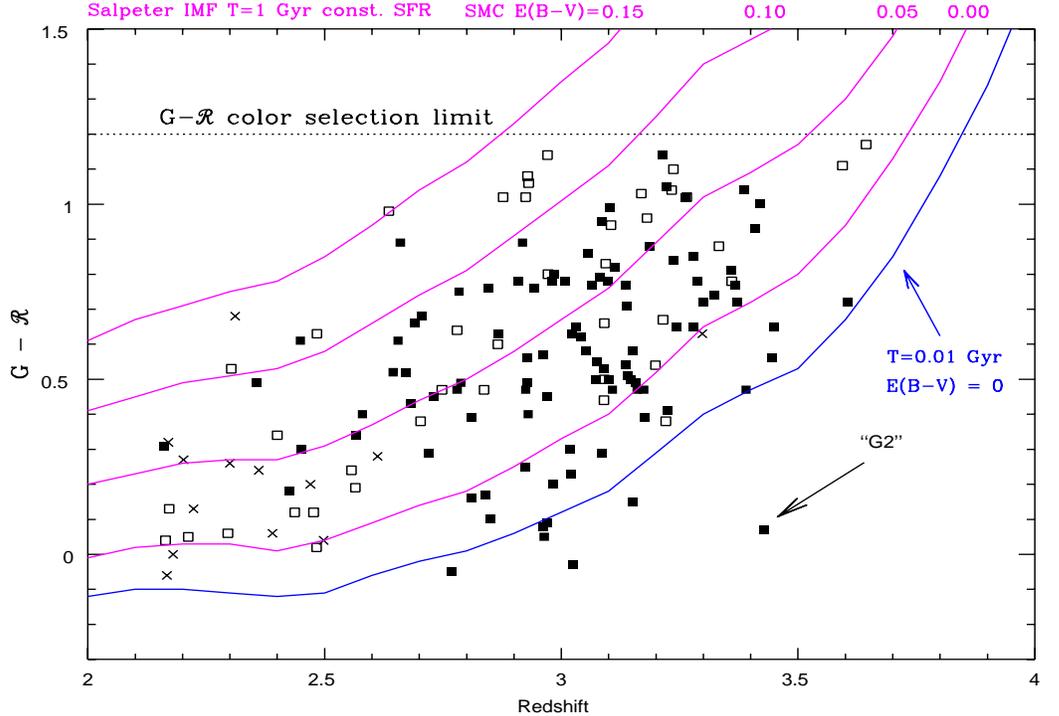,width=15.0cm,height=10.0cm,angle=270}
\hspace*{-0.75in}
\caption{Comparison between observed and predicted 
($G-\cal{R}$) colours of Lyman break galaxies 
for different amounts of SMC-type ultraviolet extinction and for 
different ages of the stellar populations. 
The symbols correspond to 
different photometric selection criteria, but all the galaxies plotted here 
have spectroscopically confirmed redshifts.}
\end{figure}

The difference between the observed and predicted ($G-\cal{R}$) colour
yields the extinction at 1500~\AA, $A_{1500}$,
appropriate to each galaxy; by adding together the individual 
values for all the galaxies in the sample
Dickinson et al. (1997)  
deduce a net correction by a factor of 1.8
to the comoving volume-averaged star formation rate 
(in \Ms~yr$^{-1}$~Mpc$^{-3}$)
at $z = 3$\,. 
Using the greyer extinction law deduced by 
Calzetti, Kinney, \& Storchi-Bergmann (1994) from
the integrated spectra of nearby starbursts 
increases the net correction to a factor of 3.5\,.
Also shown in Figure 6 is the zero-reddening curve for 
a younger stellar population (10~Myr) which 
has a bluer intrinsic UV slope $\alpha = -0.42$ (Bruzual \& Charlot 1996).
If the models are correct, some of the galaxies 
we have found are evidently younger than $10^9$ years, since they lie 
between the two zero-reddening curves in 
Figure 6.\footnote{There are a few points with ($G-\cal{R}$) colours {\it bluer}
than the $10^7$ year curve in Figure 6. 
\lya\ emission line contamination may play a role for some, particularly
for the most deviant object labelled ``G2'' which is an 
AGN with strong line emission.
Differences by $\simlt 0.1$~mag can easily be explained by photometric 
errors and the stochastic nature of the \lya\ forest 
blanketing.} Adopting the $10^7$ year model as the unreddened 
template leads to corrections to the global star formation rate
by factors of 3.5 and 6.3, for the SMC and Calzetti et al. extinction 
laws respectively.  
However, we consider it very unlikely that
{\it most} Lyman break galaxies are $\sim 10^7$ year old bursts, given that 
their number density and their masses are 
roughly comparable to those of present day 
$L^{\ast}$ galaxies (Steidel et al. 1995, 1997).

  
%
%
\begin{figure}
\vspace*{-1.5cm}
\hspace*{2.2cm}
\psfig{figure=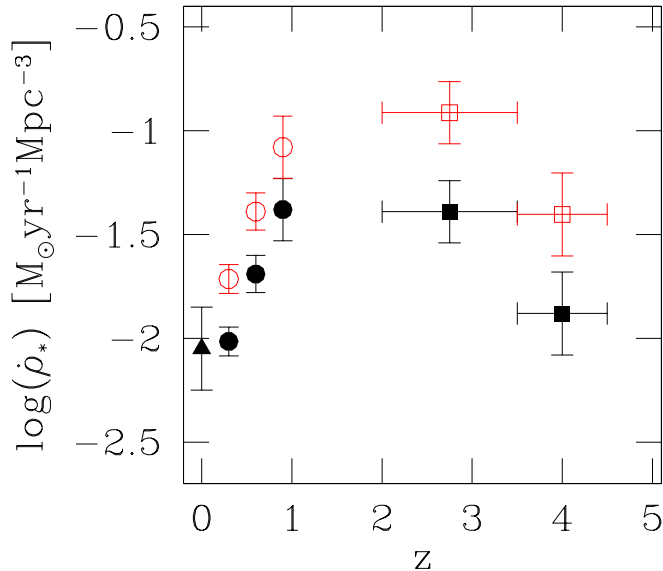,width=9.5cm,height=9.5cm,angle=0}
\hspace*{-0.75in}
\vspace*{0.5cm}
\caption{The comoving volume-averaged star formation rate as a function 
of redshift, reproduced from Madau (1997) for $H_0 = 50$~km~s$^{-1}$ and
$q_0 = 0.5$. 
The filled squares are measurements from the {\it HDF}, the circles 
from the {\it Canada-France Redshift Survey} 
and the triangle from a local \Ha\ survey. 
The open symbols show the same values corrected for dust extinction.
}
\end{figure}

We conclude that the likely dust correction to the  
integrated ultraviolet luminosity of Lyman break galaxies 
at $z = 3$ amounts to a factor of $\sim 3$. 
The correction
could be as low as $\sim 2$ and as high as $\sim 6$, 
depending on the 
age of the stellar population and on the wavelength dependence of the 
ultraviolet extinction. 
As discussed in section 5 below,
in the few cases where we have detected the redshifted
\Hb\ emission line in Lyman break 
galaxies, the line flux we measure 
confirms the relatively low values of ultraviolet extinction
deduced by Dickinson et al. (1997).


The open squares in Figure 7 show 
the effect of a factor of 3 correction for 
dust extinction on the comoving volume-averaged star formation rate 
deduced by Madau (1997)  
from the density of $U$ and $B$ drop-outs in the 
{\it Hubble Deep Field}. 
Given that the uncorrected value of $\dot{\rho}_{\ast}$ 
at any epoch deviates by only a factor of $\sim 3$ from the average
over the Hubble time, the inclusion of dust clearly has a significant impact on 
the interpretation of the cosmic star formation history.
The plot in Figure 7 is generally taken to indicate a peak 
in star formation between $z \sim 1$ and $\sim 2.5$.
This conclusion probably still holds once dust is taken into account, 
although the corrections appropriate to galaxies in the {\it 
CFRS} (Lilly et al. 1995b) have yet to be determined.
Since the values of $\dot{\rho}_{\ast}$ from the {\it CFRS} survey 
are based on galaxy luminosities in the near-UV (2800~\AA), 
the same amount of dust as we deduced for the Lyman break galaxies would 
result in an upward correction by a 
factor of $\sim 2$ (open circles in Figure 7).
A UV dust extinction of $\sim 1$~mag is supported by the very recent 
comparison between \Ha\ and UV continuum luminosities of
{\it CFRS} galaxies at $z \leq 0.3$  by Tresse \& Maddox (1997).
The revisions we propose
to Madau's plot bring the observed values of $\dot{\rho}_{\ast}$
in better agreement with recent theoretical predictions based 
on CDM models of galaxy formation
(see Figure 16 of Baugh et al. 1997).

\subsection{\lya\ Emission}

The \lya\ emission line is detected in about 75\% of the galaxies in our 
sample but is always weaker than expected on the basis of the UV 
continuum luminosities, in agreement with the generally null results of 
previous searches for high redshift galaxies based on this spectral feature
(e.g. Thompson et al. 1995).
There are strong indications that the main reason for the weakness of 
\lya\ emission is resonant scattering in an outflowing interstellar medium.
When detected, the emission line is generally redshifted by up to several 
hundred km~s$^{-1}$ relative to the interstellar absorption lines and, in 
the best observed cases, its profile is clearly 
asymmetric.

The bright galaxy 0000-D6, reproduced in Figure 8, is a good example.
The zero point of the velocity scale in Figure 8 is at 
$z = 2.960$, the redshift of 
the strong interstellar absorption lines in 0000-D6. 
On this scale the peak of \lya\ emission is 
at a relative velocity of 800 km~s$^{-1}$, and while the red wing 
extends to $\sim 1500$~km~s$^{-1}$, the blue wing is sharply 
absorbed. This P~Cygni-type profile can be understood as originating in an 
expanding envelope around the H~{\sc II} region; 
the unabsorbed \lya\ photons we see are those
back-scattered from the receding part of the nebula. 
In agreement with this picture, we find that the
systemic velocity of 
the star-forming region is $\approx 400$~km~s$^{-1}$,
as measured from the wavelengths of 
weak photospheric lines from O stars 
(S~{\sc V}~$\lambda 1501.96$ and O~{\sc IV}~$\lambda 1343.35$),
which can be discerned in this high S/N spectrum, and of
[O~{\sc III}]~$\lambda5007$ which 
we have detected in the $K$-band (see \S5 below).
Taken together, the relative velocities of interstellar, stellar and 
nebular lines point to large scale outflows in the interstellar medium, 
presumably as a consequence of the starburst activity
which in this galaxy, one of the brightest in our sample, 
approaches $\approx 100~h_{70}^{-2}$~\Ms~yr$^{-1}$.
Similar, although generally less energetic, outflows are seen in local 
starburst galaxies observed with {\it HST} and {\it HUT} 
(Kunth et al. 1996; Gonz\'{a}lez Delgado et al. 1997).

%
%
\begin{figure}[t]
\unitlength1cm
\begin{picture}(16.5,16.5)  
{\epsfxsize=16.5cm
\epsfbox{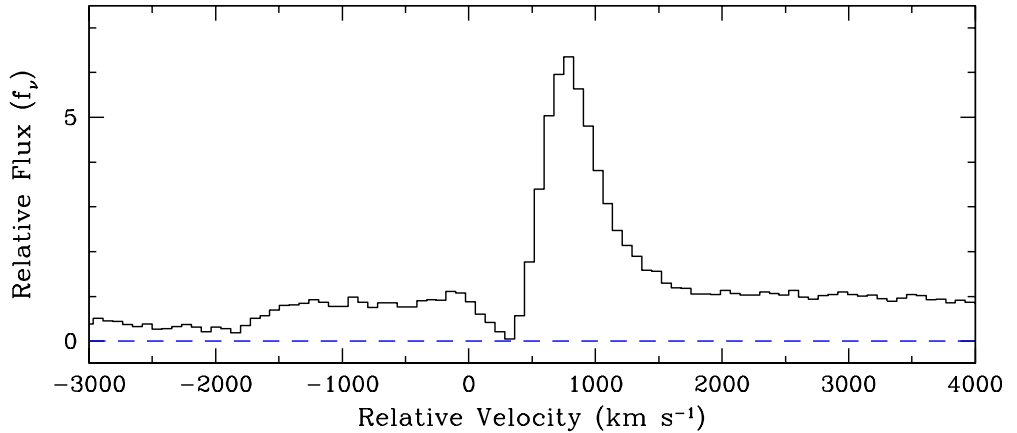}}
\end{picture}
\vskip -10cm
\caption{The wavelength region near \lya\ in the
${\cal R} = 22.9$, 
$z = 2.963$ galaxy 0000-D6.
The spectrum was obtained with LRIS on Keck I by Hy Spinrad and Arjun Dey
with an  
exposure time of 17\,650~s and a resolution of
4~\AA\ FWHM. 
The equivalent width of the combined \lya\ emission and absorption
feature is 10.5 \AA. The velocity scale is relative to the redshift of 
the strongest interstellar absorption lines.
}
\end{figure}

Large scale motions of the type we have found in 0000-D6 could be the 
main reason for the strengths of the interstellar absorption lines in 
high-$z$ galaxies which, with typical equivalent widths of 
2-3~\AA, are often greater than their counterparts in nearby starbursts.
(Since these lines are saturated, their equivalent widths are much more 
sensitive to the velocity dispersion of the gas than to the metallicity 
of the gas).
On the other hand, such strong absorption lines are also often
seen in damped \lya\ systems, which are not generally
associated with sites of active star formation and where they may
reflect the complex velocity fields of merging protogalactic clumps
(Haehnelt, Steinmetz, \& Rauch 1997).

In any case, the systemic redshift of \lya\ emission relative to 
interstellar absorption {\it along the same sight-line} brings into 
question the validity of interpreting such differences along adjacent 
sight-lines as evidence for large rotating disks, as recently proposed by
Djorgovski et al. (1996), 
and Lu, Sargent, \& Barlow (1997).

\section{Infrared Prospects}

At $z \sim 3$ the
nebular emission lines which dominate the optical spectra 
of star forming galaxies are redshifted 
into the infrared $H$ and $K$ bands; there is a strong incentive to 
detect and measure these lines as they 
hold important clues to the nature of the population of high-$z$ galaxies
we have isolated.
In  particular: {\it (a)} the line widths, which presumably reflect the 
overall kinematics of the star forming regions in a galaxy,  
can provide an indication of the masses involved; 
{\it (b)} a detection of \Hb\ (or \Ha\ at $z \simlt 2.5$) 
would give a measure of the star formation rate which can be compared 
with that deduced from the UV continuum; and 
{\it (c)} the ratios of the familiar nebular lines are probably the 
most promising way of estimating the metallicity of these galaxies, 
given the complexity of the ultraviolet absorption line spectra.

Somewhat paradoxically (given that the discovery of $z \sim 3$
galaxies awaited the availability of large telescopes)
the detection of the strongest rest-frame 
optical emission lines in the $K$-band is 
actually within reach of 4-m telescopes equipped with
moderately high dispersion near-infrared spectrographs.
What is required is prior knowledge of the galaxy redshift and sufficient 
spectral resolution to ensure that the lines of interest fall in 
a gap between the much stronger OH$^-$ emission features which dominate 
the infrared sky.
Pilot observations which we carried out with the CGS4 spectrograph on UKIRT 
in September 1996 were successful in detecting \Hb\ and/or [O~{\sc III}] 
emission lines in both $z \simeq 3$ galaxies targeted, 
0000-D6 and 0201-C6 \cite{pettini97a}; 
a third measurement---in cB58---has been obtained by 
Gillian Wright (private communication).
Figure 9 shows a portion of the $K$-band spectrum of  
0201-C6. \Hb\ and the weaker member [O~{\sc III}] doublet are both detected at 
the $\sim 5 \sigma$ level, 
whereas the stronger [O~{\sc III}] line is lost in the 
nearby sky emission.

%
%
\begin{figure}[t]
\unitlength1cm
\begin{picture}(16.5,16.5)  
{\epsfxsize=16.5cm
\epsfbox{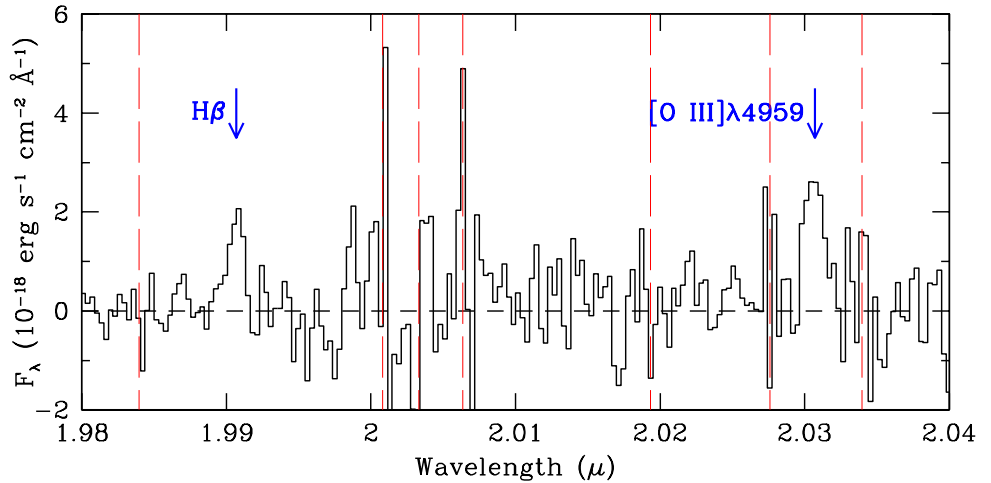}}
\end{picture}
\vskip -9cm
\caption{Portion of the infrared spectrum of the ${\cal R} = 23.9$, 
$z = 3.059$ galaxy 0201-C6 obtained with CGS4 on UKIRT
in September 1996. 
The exposure time was 18\,000~s and the resolution is
8~\AA\ FWHM sampled with 2.7 wavelength bins, each 3~\AA\ wide.
The vertical dashed lines indicate the locations of the
major OH$^-$ sky emission lines.
}
\end{figure}

As can be seen from Figure 9, both \Hb\ and [O~{\sc III}]
emission lines are resolved; after correcting for the instrumental 
resolution, we measure $\sigma = 70 \pm 20$~km~s$^{-1}$\,.
A similar velocity dispersion is also found in the other two cases, 
0000-D6 and cB58. 
(Incidentally, the fact that cB58 shows the same velocity dispersion 
as the other two $z \simeq 3$ galaxies, even though it is $\simgt 10$ 
times brighter, is another indication of its gravitationally lensed 
nature). 
If we combine $\sigma = 70$~km~s$^{-1}$ with the half-light radii
of $\approx 2$~kpc deduced for both 
0201-C6 and 0000-D6 from {\it HST} WFPC2 images, 
we obtain virial masses
of $\sim 1.2  \times 10^{10}$\Ms. 
This is comparable to the mass of the Milky Way bulge (Dwek et al. 1995)
and to the dynamical mass within the central $r = 2$~kpc 
of an $L^{\ast}$ elliptical galaxy. However, 
the total masses involved are likely to be substantially greater,
given that the present 
IR observations sample only the innermost cores of the galaxies, 
where the star formation rates are presumably highest.
As we discussed in \S3 above,  the clustering properties of 
Lyman break galaxies strongly suggest that they are associated 
with dark matter halos of mass $M \simgt 10^{12}$\Ms.

The \Hb\ flux of 0201-C6 (Figure 9), 
$(2.6 \pm 0.6) \times 10^{-17}$ erg~s$^{-1}$~cm$^{-2}$,
corresponds to a  luminosity
$L_{\rm {H}\beta} = (2.3 \pm 0.5 ) \times 10^{42}~h_{70}^{-2}$ erg~s$^{-1}$.
Adopting an \Ha/\Hb\ ratio of 2.75
and Kennicutt's (1983)
calibration 
{\it SFR} (\Ms yr$^{-1}$) 
$= L_{\rm {H}\alpha}$ (erg~s$^{-1}$)/$1.12 \times 10^{41}$ 
which is appropriate for a Salpeter IMF from 0.1 to 100 \Ms, 
we deduce a star formation rate 
{\it SFR}$_{\rm {H}\beta} = (55 \pm 13)~h_{70}^{-2}$~\Ms\ yr$^{-1}$.
For comparison 
{\it SFR}$_{\rm{UV}} = (20 - 35)~h_{70}^{-2}$~\Ms yr$^{-1}$, 
depending on whether the age of the starburst is 
$10^9$ or $10^7$ years respectively.
Estimates of the star formation rate from \Ha\ emission and the UV continuum do 
not normally agree to better than a factor of $\sim 2$ in local starbursts
(e.g. Meurer et al. 1995); the agreement in 0201-C6 is further improved when 
account is taken of the small amount of reddening ($E(B-V) \simlt 0.1$)
implied by the slope of the UV continuum ($\alpha = 0.35$) using the 
prescription by Calzetti (1997a).
We reach similar conclusions in 0000-D6 and cB58; in all three cases 
ultraviolet extinctions by factors of $\approx 2-3$ are indicated by the 
comparison between  
UV and \Hb\ luminosities.

 
These preliminary results demonstrate clearly the great potential of 
the infrared spectral region for complementing the information provided by 
the rest-frame ultraviolet and ultimately 
leading to a better understanding of the 
nature of high-$z$ galaxies. 
With large telescopes
the detection of nebular emission lines in the near-IR will be
easier and it will be possible to address the points touched upon 
here in greater depth, using a large set of measurements.
On the other hand, the strong sky background will remain a fundamental 
limitation of ground-based observations and will be fully overcome 
only with space observatories such as the {\it NGST}. 
There is therefore a strong motivation to provide {\it NGST} with 
at least a moderately high spectral resolution capability 
($R \geq 3000$), adequate to resolve the emission lines from 
star forming galaxies.
Free from the complications introduced by the Earth's atmosphere,
{\it NGST} will be able to record the full complement of nebular emission 
lines and obtain reliable estimates of the abundances of several elements.
By bringing together abundance measurements
from emission and absorption 
(Lu et al. 1996; Pettini et al. 1997) line data
it will be possible to build a full picture
of the early stages in the chemical enrichment history of the universe.

\section{Epilogue}

Star forming galaxies with spectra broadly similar to those of present day 
starbursts have now been detected up to redshift $z \simeq 5$ (Franx et 
al. 1997). It is remarkable that at such an early epoch, 
only 1.5~Gyr after the Big Bang, many of the characteristics of the 
present day universe were already in place. The intergalactic medium was 
fully ionised, galaxies were enriched in a wide variety of 
chemical elements (albeit in lower proportions than today),
and structures on the scale of rich clusters 
were already evident.
As we learn more about the time when the luminous galaxies of today
were beginning to form, 
already the next goal in our {\it `ORIGINS'} quest is coming 
into focus: identifying the time when the first stars 
formed, produced the first generation of heavy elements and ionised the 
universe which had remained neutral since the cosmic microwave background 
was emitted at $z = 1000$.

Our first steps in this unfamiliar territory are guided, as it is often the 
case, by theoretical considerations. The lowest metallicities 
measured at $z = 4 - 5$ are $\sim 1/300$ of solar (Pettini et al. 1995; 
Lu et al. 1996). As discussed by Madau \& Shull (1996) and 
Miralda-Escud\'{e} \& Rees (1997), the ionising photons associated with 
the production of even such trace amounts of heavy elements outnumber the 
baryons by 10 to 1 (this is a rather straightforward calculation, because 
the same massive stars are the source of heavy elements and ionising 
photons). Thus, if this initial enrichment took place when the universe 
was mainly neutral, the associated Lyman continuum radiation was 
sufficient to reionise all the diffuse gas.

Recently Miralda-Escud\'{e} \& Rees (1997) 
and Rees \& Miralda-Escud\'{e} (1997)
have explored some of the consequences of this scenario.
They point out that the physics of the galaxy formation process is different
depending on whether galaxies form by accreting neutral or photoionised 
gas, mainly because the cooling efficiency is different in the two cases.
They speculate that the first objects to collapse during the `dark ages' 
of the universe are likely to be on very small scales, 
with velocity dispersions
$\sigma \simlt 10$~km~s$^{-1}$ and masses $M_b \simlt 10^8$~\Ms.
Under these circumstances the brightest sources at very high redshifts
are likely to be Type II supernovae.  
At $z = 10$ such Population III supernovae
may reach an infrared magnitude 
$K \simeq 30$ and may be as numerous as one supernova 
per square arcminute per year. 
While it is just conceivable that an event of this kind may be  
within reach of current instrumentation if gravitationally
lensed by a foreground cluster of galaxies, testing these bold new ideas
will be the mission of the {\it Next Generation Space Telescope}.

\end{document}